# Evolution of seismic signals and slip patterns along subduction zones: insights from a friction lab scale experiment.

Christophe Voisin, Jean-Robert Grasso, Eric Larose, François Renard[*]

*Laboratoire de Géophysique Interne et Tectonophysique, CNRS, Université Joseph Fourier, Maison des Géosciences, BP 53, 38041 Grenoble cedex 9, FRANCE.*

[*]also at Physics of Geological Processes, Oslo, Norway

**Continuous GPS and broadband seismic monitoring have revealed a variety of disparate slip patterns especially in shallow dipping subduction zones, among which regular earthquakes, slow slip events and silent quakes[1,2]. Slow slip events are sometimes accompanied by Non Volcanic Tremors (NVT), which origin remains unclear[3], either related to fluid migration or to friction. The present understanding of the whole menagerie of slip patterns is based upon numerical simulations imposing *ad hoc* values of the rate and state parameters $a$ and $b$[4-6] derived from the temperature dependence of $a$ and $b$ of a wet granite gouge[7]. Here we investigate the influence of the cumulative slip on the frictional and acoustic patterns of a lab scale subduction zone. Shallow loud earthquakes (stick-slip events), medium depth slow, deeper silent quakes (smooth sliding oscillations) and deepest steady-state creep (continuous sliding) are reproduced by the ageing of contact interface with cumulative displacement[8]. The Acoustic Emission evolves with cumulative displacement and interface ageing, following a trend from strong impulsive events, similar to earthquake seismic signals, to a collection of smaller amplitude and longer duration signals, similar to Non Volcanic Tremors. NVT emerge as the recollection of the local unstable behaviour of the contact interface globally evolving towards the stable sliding regime.**

Shallow dipping subduction zones are the best natural laboratories to observe and measure the whole gamut of slip patterns: Large seismic events break the upper locked portion of the subduction plate into events of magnitude up to 9.3-9.4 so far; M=6.0-7.5 slow slip events (SSE) or silent events occur down-dip the locked portion of the subduction zone[1,2,9]. Deeper along these subduction zones, the oceanic plate slips continuously at the plate tectonics velocity (Figure 1). SSE and silent events are sometimes accompanied by Non Volcanic Tremors (NVT)[2]. A tremor represents long duration of high-amplitude seismic signals without clear body wave arrivals[10].

The physical origin of NVT is poorly understood, being either proposed to be triggered by fluid movements[11] or related to some frictional processes along the subduction interface[12,13]. Both origins are problematic. If fluids are at the origin of tremors, then we should record tremors up-dip of the locked zone, where the subducted lithosphere is known to be fluid-saturated. If tremors do find their origin in some frictional process, then we should record NVT not only down-dip the locked zone, but all along the subduction zone. Whatever the process at the origin of NVT, the fact is that they are mostly associated with SSE along subduction zones, suggesting a common origin to both phenomenons. The understanding of SSE is incomplete. Pressure and temperature (P & T) increases with depth are invoked to explain the changes of slip patterns down dip the subduction zone[10]. Aqueous fluids released from subsequent metamorphic dehydration reactions (enhanced by P&T changes) have been proposed to play an important role in causing slip transients and NVT[14]. From the numerical point of view, P&T changes are used to establish the profile of Rate and State friction parameters *a* and *b* with depth[5,6]. *b-a* is tuned to control the stable/unstable behaviour of the subduction interface, being positive or negative respectively. The depth of the transition zone is set up according to the temperature dependence of *b-a*. This dependence was experimentally determined for a wet granite gouge[7]. Other studies impose small but negative (*b-a*) values at the transition zone to reproduce the change in slip pattern. In this latter case, the conditional stability of slip is insured by either a stiffness that is greater than the critical one[15] or by a plastic increase of the critical slip distance $d_c$[4]. In all of these models, the subduction interface is imposed to slip at a constant tectonic velocity below the transition zone. Numerical models of friction applied to subduction zones teach us that the transition between earthquakes and SSE occurs down dip the locked zone, where the slip becomes conditionally stable. However, the same situation for stability exists also up-dip of the locked zone[16] where no SSE has been observed so far. The process(es) at the origin of SSE must explain this difference in the slip pattern above and below the locked zone, respectively. The occurrence of NVT is not investigated by these numerical models.

In the following, we investigate the influence of cumulative slip and ageing of the contact interface on the occurrence of SSE and NVT. The cumulative slip along subduction zones is among the largest on Earth: indeed, a portion of the subducted lithosphere slips over hundreds of kilometres. During this slip, two main processes drive the change in the contact interface: 1/ the sediment input in the subduction factory that constitutes a highly deformable contact interface; 2 / P & T increases, known to drive changes in rock deformation from elastic brittle to plastic creep behaviours. We have devised a friction experiment (Figure 1) that is aimed at reproducing the main features of a natural subduction zone. To recover P & T effects on rock deformation, we allow for plastic deformation to be effective over the time scale of our friction experiment by using a salt slider[8, 17]. The slider, an analogue to a portion of the subduction plate, slides against a flat glass surface which represents the overriding continental lithosphere. The slider is pushed at a constant velocity, equivalent to the tectonic velocity. We follow the displacement of the slider through time, as if we were moving along with the buried lithosphere (Figure 1). We record the Acoustic Emission (AE) induced by the slip of the slider.

The slip pattern of the slider is observed to evolve continuously with the cumulative displacement. Figure 2 presents four exerts of this continuous change, taken for different stages of cumulative slip. We observe a change in the frictional behaviour of the slider, from unstable (stick-slip like) to stable slip, passing through quasi stable smooth oscillations. The beginning of the experiment is characterized by large stick-slip events (Figure 2, snapshot 1) which amplitude gradually decreases as the slider enters a smooth oscillations sliding regime (snapshots 2-3). The slip pattern of the slider eventually reaches the continuous sliding regime at the imposed driving velocity (snapshot 4). If the amplitude of slip events is decreasing with the cumulative displacement, the duration of slip episodes increases with the cumulative slip. The stick-slip episodes are characterized by a fast stress release (< 0.05 second, limited by the sampling frequency), while the duration of the smooth sliding oscillations is about 30 seconds.

In a previous study, we reported on the change in the contact interface geometry with cumulative displacement[8]. The change in slip pattern of the slider is accompanied by the development of a striation pattern in the direction of sliding. Here we record the AE associated with the slider slippage and track any change in the acoustic recordings in relation with the change in the contact interface. Figure 3 presents three AE records taken after different increasing cumulative slips. The amplitude of AE signals is decreasing with the cumulative displacement. On the contrary the duration of AE signals increases with the cumulative displacement. The envelope of AE signals is evolving as well. The three examples show a transition from strong impulsive signals to a succession of smaller and longer signals, made of distinct acoustic events. Acoustics of friction at the scale of 1 square centimetre is generally related to the surface roughness[18]. Recording the AE at high frequency during the friction experiment provides a way to resolve the processes of ageing going on at a lower scale of friction: the contact asperity scale. In our experiments, the progressive evolution of the acoustic signals is related to the contact interface changes and the development of the striation pattern. The records of AE throughout the experiments reveal a progressive change from impulsive events to smaller and longer signals made of numerous acoustic events. It reproduces the in-situ change from earthquake seismic signals to NVT signals observed in subduction zones. Our experiments argue for slow slip events

(SSE) and non volcanic tremors (NVT) being two manifestations of a single friction process, seen at two different scales. These experimental results shed a new light on both the slip and seismic patterns along the subduction zones. In our experiments, tremor like signals are emitted by unstable slip at the contact asperity scale (shear failure asperities) occurring during the slider slow slip event. Ageing of the contact interface with cumulated displacement do occur along the subduction interface, driving the change from regular to slow to silent earthquakes. Subduction zone NVT appear as a local recollection of the unstable frictional behaviour occurring during the slippage of aseismic slow or silent events down dip the subduction zones[19]. The occurrence of slow and silent events, possibly twined to NVT is causally linked to the amount of cumulated slip undergone by the deformable contact interface. This explains why neither SSE nor NVT are recorded up-dip of the locked zone, where the accumulated slip is too small.

The different classes of slip events (earthquakes, slow events, silent events, continuous sliding) are based on observations, either seismic or geodetic. To press further the comparison between our experimental results and the subduction zone events, we estimate slip velocity ratios for the four classes of events (Figure 2). The normalized slip velocity V* ($V_{event}$ / $V_{driving}$) is different for subduction zones and for the analogue experiment, as far as fast unstable slips are concerned. Because of a low sampling frequency, we cannot be conclusive on this apparent misfit. Nonetheless, V* scales the same way for subduction and analogue events. Our experimental results suggest that earthquakes, slow and silent events pertain to a continuum of slip patterns driven by the ageing of the subduction interface with cumulated slip. Estimates of the cumulated slip required to observe slow or silent events on actual subduction zones are given by the depth of these events and the dip angles of subduction zones. Compiling data from Kato [2003] we derive a value for cumulated slip of 70 to 230 km, with a mean value of 130 km. Using the appropriated tectonic plate velocities[20], the necessary time for these slow and silent slip events to emerge is in the 1.5-4 $10^6$ yr range, with a 3 $10^6$ yr mean value.

Allowing deformation of the contact interface to interfere with friction reveals a continuum of slip patterns. Earthquakes, slow events, silent quakes and continuous sliding appear as different aspects of a holistic process. The ageing of the contact interface with cumulated displacement provides a global framework to capture the occurrence of the different slip patterns and seismic signals along subduction zones. Considering the cumulative displacement as a *sine qua non* condition for the occurrence of SSE and NVT explains the absence of these latter above the locked zone. Our experimental results are consistent with and rationalize *a posteriori*: (i) the modelling of aseismic slip transients by a decrease in *b-a*[5] and an increase in $d_c$[4]; (ii) the hypothesis that silent slip and NVT pertain to one and unique phenomenon of friction; (iii) the hypothesis that NVT are local reminiscence of frictional instabilities in these aseismic slip transients[13].

Correspondence and requests for data should be addressed to CV (cvoisin@obs.ujf-grenoble.fr)

**Figure 1. (A)** Scheme of a subduction zone geometry with the relative location of large earthquakes (EQ), slow and silent events (SSE), and continuous stable sliding (CS). The oceanic plate (pink) is moving at constant imposed velocity (the tectonic loading rate) and is subducted under a continental lithospheric plate (yellow). The experimental analogue to the subduction zone is a salt slider block (light grey) pushed at constant velocity over a planar glass surface. Pressure and temperature effects are mechanically introduced by allowing creep deformation to be efficient at the time scale of the friction experiment. Different frictional slip patterns of the slider are observed: Stick slip (SS) followed by quasi-stable smooth oscillations (SO) and by continuous sliding (CS). **(B)** Overview of the friction experiment. The salt sample (1 cm$^2$ surface area) is housed in a nickel iron alloy (Invar©) plate with a low thermal expansion coefficient to minimize thermal perturbations. The salt slider is carefully sliced from a single crystal of NaCl to avoid crack formation. The frictional interface is then roughened with a sand paper (Struers #320 grit) to ensure an initial roughness. A constant continuous velocity of 0.11 µm.s$^{-1}$ is imposed to the plate. The slider is subjected to a constant normal load of 0.24 MPa and is in contact with a flat glass surface. Five displacement transducers ($d_{1-5}$) record the plate movement in horizontal ($d_{1-2}$) and vertical ($d_{3-5}$) directions. A force gauge records the shear force exerted onto the slider. An accelerometer (reference #4518 from Bruel & Kjaer) is glued on the upper surface of the slider and records the acoustic emission in the 20 Hz-70 kHz frequency range.

.

**Figure 2.** The continuous variation of the slip pattern of the salt slider. Upper line, from left to right: regular stick slip, smooth stick slip, smooth oscillation, continuous sliding. Lower line: the normalized slip velocity V* in the four cases (V*=V/V$_{driving}$). V$_{driving}$ is cm/year for subduction zones, and m/year for the analogue experiment. In the experiment, V* is about 300, 60, 10, 1, for stick slip, smooth stick slip, smooth oscillations and continuous sliding respectively. For subduction zones, with typical velocities of m/s, cm/s and µm/s, V* is about $10^9$, $10^7$, 10, 1 for earthquakes, slow events, silent events, continuous sliding respectively. The duration of each event is defined by the time period during which the slip velocity is two times greater than the driving velocity. This duration is lower than 0.05 s for the regular stick slip; about 5 s for smooth stick slip; 25 s for smooth oscillations; not defined for continuous sliding.

**Figure 3.** Three examples of Acoustic Emission records taken after different increasing cumulative slips (0.5 cm; 1.5 cm; 2.5 cm respectively). The horizontal scale is the time in seconds. The time t=0s is the triggering time. The vertical scale represents the vertical acceleration of the slider, given in m.s$^{-2}$. The amplitude of AE signals is decreasing with the cumulative displacement (2, 1, 0.05 m.s$^{-2}$ respectively). On the contrary the duration of AE signals increases with the cumulative displacement (~$10^{-4}$s; ~$10^{-3}$s; ~$10^{-2}$s respectively). The envelope of AE signals is evolving as well. The three examples show a transition from strong impulsive signals to a succession of smaller and longer signals, made of distinct acoustic events. Note that no AE could be recorded during the continuous sliding stage because of the triggered mode of recording.

**Figure 4**. The change in the Acoustic Emission of the slider. Three consecutive experiments increase the cumulative slip by 1 cm each: PA006 (0 to 1cm); PA007 (from 1 to 2cm); PA008 (from 2 to 3cm). For each experiment, all acoustic events are recorded and stacked in order to increase the signal to noise ratio. The results are presented in the form of a spectrogram. The low frequency range of the signals recorded below 10 kHz is associated with the movement of the slider as a block. The high frequency range (around 50 kHz) is associated with the fast slip of the contact asperities. The time t=0s is the triggering time. The time window is focused between -0.02s and +0.02s. Some differences appear in the form of signals recorded. During the first centimetre of slip, the high frequency signal is recorded as a single large acoustic event. During the second centimetre of slip, the acoustic signals have smaller amplitudes but remain in the form of a unique acoustic event. Some weak signals begin to appear in the 0-0.01s time window. During the third centimetre of slip, the amplitude of the records is weaker than previously recorded. The high frequency part of signal is made of a collection of different events occurring at various times, before and after the triggering time.

Table 1. A comparison of slip ratio, slip duration ratio and slip velocity ratio (V*) between earthquake and slow or silent slip events for the subduction zone and for the analogue experiment. Slip ratio is $slip_{earthquake}$ / $slip_{event}$; Slip duration ratio is $duration_{earthquake}$ / $duration_{event}$; Slip velocity ratio is given by V* = $V_{event}$ / $V_{driving}$. $V_{driving}$ is cm/year for subduction zones, and m/year for the analogue experiment. $V_{event}$ is defined as the ratio of slip over duration. The low temporal resolution of the experiment (20 Hz) precludes to explore high speed velocities, which may explain the discrepancies between V* for fast slips. In spite of this, the concordance of all ratios between the subduction zone and the analogue experiment supports the control of slip pattern by the cumulative displacement and the deformation of the contact interface.

**Figure 1**

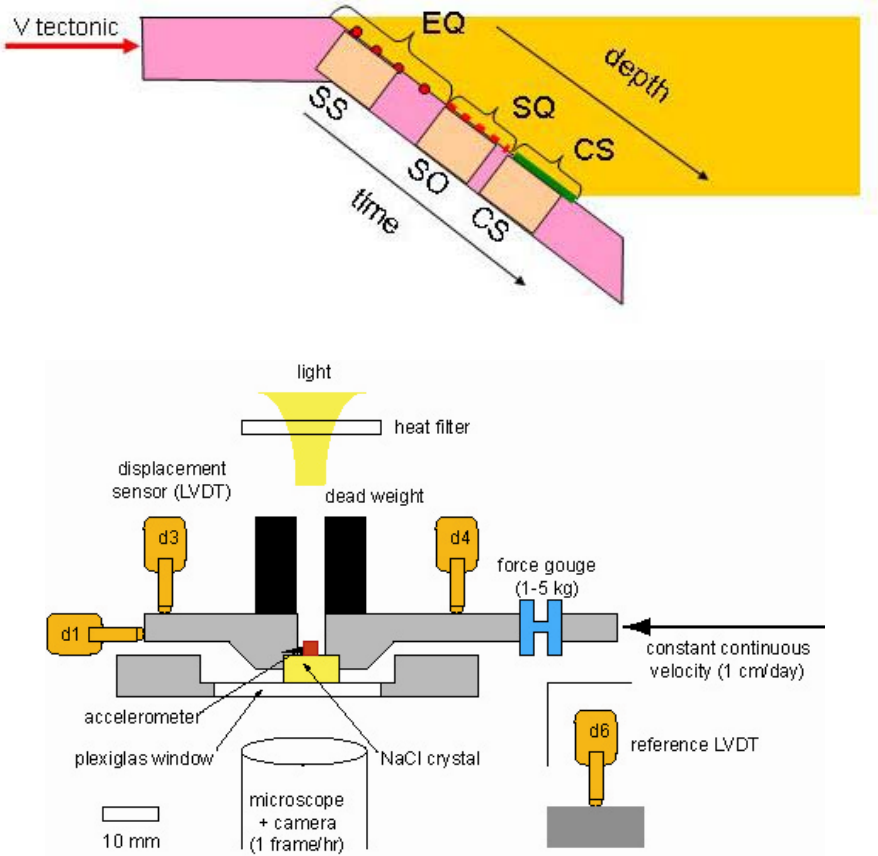

**Figure 2**

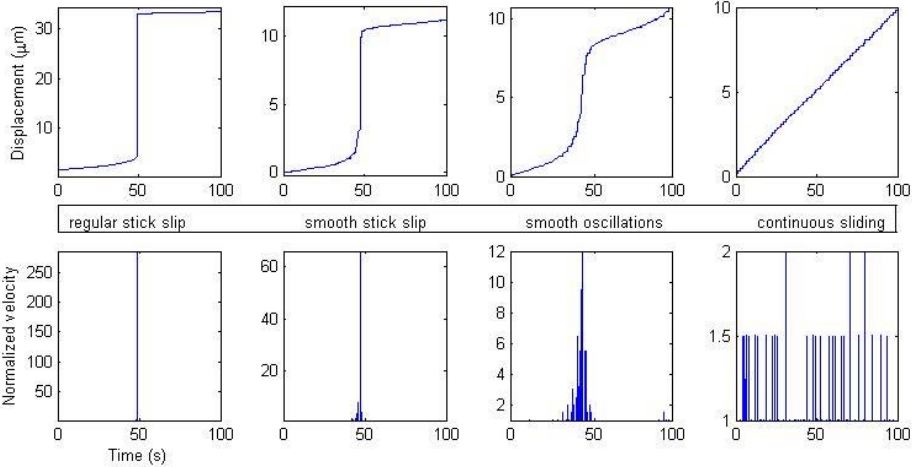

**Figure 3**

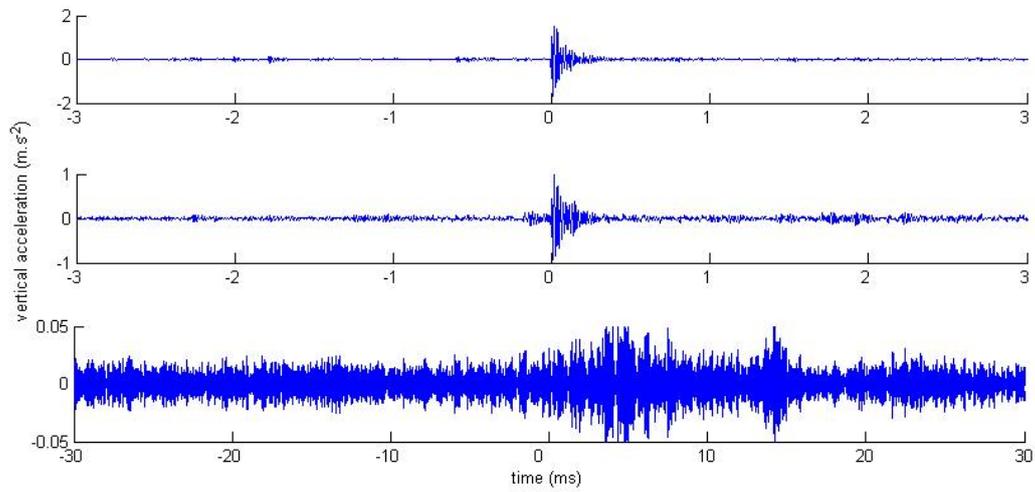

**Figure 4**

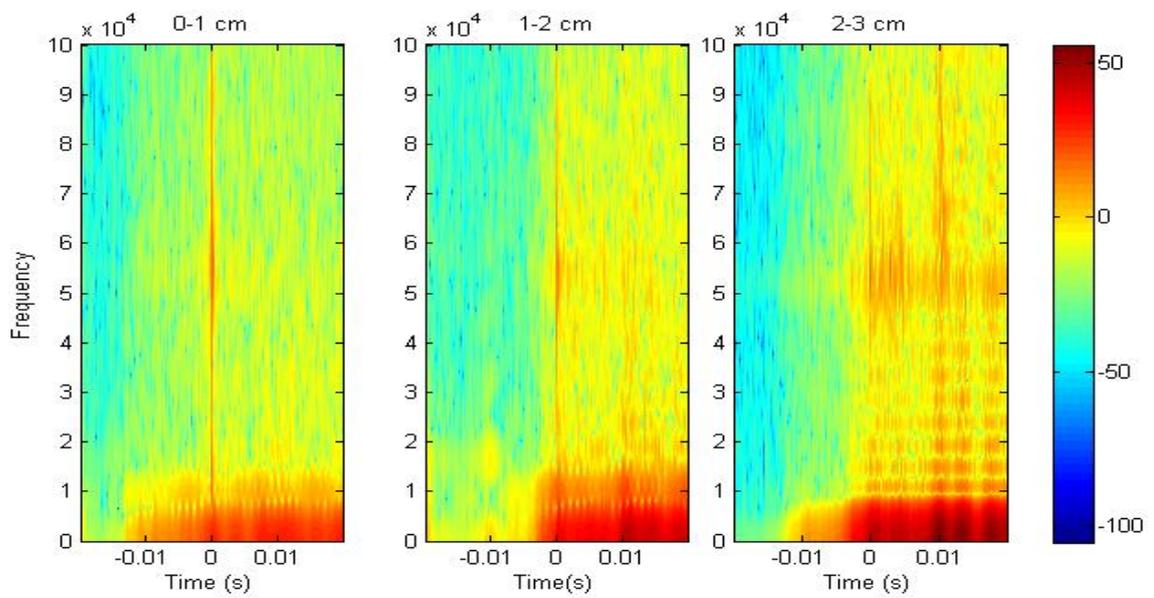

**Table 1**

| | subduction zone | | | analogue experiment | | |
|---|---|---|---|---|---|---|
| slip | earthquake ~1 m | slow event ~0.5 m | silent quake ~0.1 m | stick slip ~30 μm | smooth stick slip ~10 μm | smooth oscillation ~5 μm |
| slip ratio | 1 | 2 | 10 | 1 | 3 | 6 |
| duration | 10 s | hours-days | days-years | <0.05 s | ~5 s | ~25 s |
| duration ratio | 1 | $3.10^{-3} - 10^{-4}$ | $10^{-3} - 10^{-5}$ | 1 | $< 10^{-2}$ | $< 2.10^{-3}$ |
| $V^*$ | earthquake ~$10^9$ | slow event ~$10^7$ | silent quake ~2-10 | stick-slip $6.10^3$ | smooth stick-slip ~20 | smooth oscillation ~2 |